\documentclass[aps,prl,twocolumn,floatfix]{revtex4}

\usepackage[pdftex]{graphicx}
\usepackage{mathtools,amsthm,amssymb}
\usepackage[pdfauthor={Mathias Diez and Guido Burkard},pdftitle={D'yakonov-Perel'-type spin relaxation in bilayer graphene},pdfsubject={},colorlinks=true,linkcolor=blue,citecolor=magenta]{hyperref}
\usepackage{dcolumn}
\usepackage{upgreek}
\usepackage{bbm}
\usepackage{overpic} %%%%MD

\begin{document}

\title{Bias-dependent D'yakonov-Perel' spin relaxation in bilayer graphene}
\author{Mathias Diez and Guido Burkard}
\address{Department of Physics, University of Konstanz, D-78457 Konstanz, Germany}
\date{\today}
\begin{abstract}
We calculate the spin relaxation time of mobile electrons due to 
spin precession between random impurity scattering (D'yakonov-Perel'
mechanism) in
electrically gated bilayer graphene analytically and numerically.  
Due to the trigonal warping of the bandstructure, the spin relaxation
time exhibits an interesting non-monotonic behavior as a function
of both the Fermi energy and the interlayer bias potential.
Our results are in good agreement with recent four-probe measurements 
of the spin relaxation time in bilayer graphene and indicate the
possibility of an electrically-switched spin device.
\end{abstract}

\maketitle

Many fascinating properties of electrons in graphene have been brought
to light since its discovery, such as their high electron mobility 
and the emergence of anomalous integer 
quantum Hall plateaus \cite{Novoselov2005,Zhang2005,CastroNeto2009}.
One of the less studied but important questions 
is the capability of graphene to store and transport electron spin.
Compared with semiconductors such as Si or the III-V compounds, 
graphene bears superior traits for long spin coherence:
its low density of nuclear spins reduces hyperfine interactions that
are limiting spin coherence in GaAs, while its low atomic weight
implies intrinsically weak spin-orbit interactions (SOI) thus
allowing for slow spin relaxation \cite{Trauzettel2007}.

Graphene spin valve devices have been demonstrated soon after
the discovery of graphene \cite{Hill2006}, followed by four-probe 
spin transport experiments using ferromagnetic cobalt \cite{Tombros2007} and
permalloy \cite{Cho2007} electrodes.  From Hanle precession 
measurements, spin relaxation times on the 
order of 150 ps were found, and by simultaneously modifying
the mobility and spin relaxation time by tuning the Fermi energy
with an external gate, a behavior of the spin relaxation time
consistent with an Elliot-Yafet type mechanism was identified \cite{Tombros2007}.
Subsequent experiments have confirmed this for single-layer 
graphene \cite{Han2011}, but D'yakonov-Perel' type behavior
in combination with spin relaxation times up to a few nanoseconds
at 4 Kelvin were found in bilayer graphene \cite{Han2011,Yang2011}.

%%%%%%%%%%%%%%%%%%%%%%%%%%%%%%%%%%%%%%%%%%%%%%%%%%%%

Motivated by these observations, we calculate the spin relaxation rate 
for bilayer graphene according to the D'yakonov'-Perel mechanism. 
Our starting point is the band Hamiltonian of AB-stacked bilayer
graphene (BLG)
for momenta $\hbar {\bf k} = \hbar(k_x,k_y)$ 
near the Dirac points K ($\tau=1$) and 
K' ($\tau=-1$) \cite{McCann2006},
\begin{equation}
  H_\text{BLG} = \left(\begin{array}{c c c c}
     \frac{U}{2} & \tau v_3 p^* & \tau  v_F p & 0 \\
      \tau v_3 p & -\frac{U}{2} & 0 & \tau  v_F p \\
     \tau  v_F p^*  & 0 & \frac{U}{2} & \gamma_1 \\
    0 & \tau  v_F p^*  & \gamma_1 & -\frac{U}{2} \end{array}\right),
  \label{eq:H0}
\end{equation}
in the basis $A_1$, $B_2$, $B_1$, $A_2$ where $A_1$ refers to the
A-sublattice in the upper layer, $B_2$ to the B-sublattice in the
lower layer, etc., and
where $p = \hbar (k_x+i \tau k_y )  = \hbar k e^{i\tau \phi}$ with
$\phi=\arctan(k_y/k_x)$. 
Here, the intralayer hopping parameter $\gamma_0=2.8\,$eV determines the
Fermi velocity $v_F=\frac{3}{2}a\gamma_0/\hbar=8.0\cdot 10^5\,$m/s,
whereas the interlayer hopping parameter $\gamma_1=0.39\,$eV gives rise to a strong coupling of the two stacked lattice sites $B_1$ and $A_2$. 
Skew interlayer hopping with strength $\gamma_3=0.315\,{\rm eV}$ introduces an
additional velocity $v_3=\frac{3}{2}a\gamma_3/\hbar=5.9\cdot 10^4$ m/s and causes a significant trigonal warping of the energy dispersion. 
A tunable energy offset $U$ between the two layers can be achieved by
applying a bias voltage and leads to the opening of a band gap,
which has been observed to reach up to $250\,$meV \cite{Zhang2009}.
For what follows, it is important to note that the interlayer bias
also breaks inversion symmetry, and therefore, in combination with the
intrinsic SOI, can lead to a spin splitting.

The SOI in bilayer graphene is still a topic of
ongoing theoretical discussion \cite{Guinea2010,Konschuh2011}. 
The Hamiltonian of the intrinsic SOI consistent
with the crystal symmetry is found to be \cite{Guinea2010}
$H_\text{SO} = \lambda_1\tau\sigma_z s_z + \lambda_2\tau\mu_z s_z
    +\lambda_3\mu_z\left( \sigma_y s_x-\tau\sigma_x s_y \right)
    +\lambda_4\sigma_z\left( \mu_y s_x+\tau\mu_x s_y \right)$,
where $\mu_i$, $\sigma_i$ and $s_i$ are Pauli matrices denoting layer, sublattice, and electron spin, respectively. 
The last SOI parameter which is estimated to be
$\lambda_4=0.48\,$meV dominates the other terms, with
$\lambda_1=14\,\upmu$eV, $\lambda_2=8\,\upmu$eV, and 
$\lambda_3=5.5\,\upmu$eV. 
Both the $\lambda_1$ and the $\lambda_2$ terms are diagonal in spin, pseudospin
and layer leading to out-of-plane low-energy effective spin-orbit fields which do not
efficiently couple to momentum scattering as is needed for D'yakonov-Perel'-type
spin relaxation. 
The remaining two terms give rise to in-plane spin-orbit fields which change their  direction depending on the angle of the electron's momentum.
However, not only was $\lambda_3$ found to be much smaller 
than $\lambda_4$ in Ref. \cite{Guinea2010}, but in comparison with $\lambda_4$-type spin-orbit interaction the magnitude of its corresponding spin-orbit field at low Fermi energies $E_F$ is further supressed by $E_F/\gamma_1$. 
Below, we focus on the $\lambda_4$ term, for a discussion of the remaining terms of $H_\text{SO}$ including the corresponding expressions for the spin-orbit fields, see Appendix.

In the presence of SOI and for $U\neq 0$, the 
four spin-degenerate bands described by $H_\text{BLG}$ split up into
eight bands.  Half of those bands are split off from the Dirac
points by $\gamma_1$ and are not directly involved in spin
transport when the Fermi energy is in the vicinity of the 
Dirac point.  Among the remaining four low-energy bands, two 
correspond to electron and two to hole states, each with their
split spin degeneracy.  To obtain the spin-orbit field for
electrons (holes), we focus on positive (negative) Fermi 
energies, where spin currents are carried by
the electrons (holes).  In order to derive an effective model for the
low-energy bands, we perform a 
Schrieffer-Wolff transformation on the total Hamiltonian
$H=H_\text{BLG}\otimes\mathbbm{1}_S+H_\text{SO}$,
restricting ourselves to the dominant $\lambda_4$ term
for the rest of the discussion (see Appendix for a more general
discussion).  
For this purpose, we divide up the total Hamiltonian
into low- and high-energy parts (separated by $\gamma_1$),
and the interactions $V$ that couple them,
$H=H_0+V$, where $H_0$ corresponds to $H_\text{BLG}$ without
intralayer hopping ($v_F=0$), while $V$ contains both intralayer
hopping and SOI and can be expressed in the 
basis $A_{1,\uparrow}$, $A_{1,\downarrow}$, $B_{2,\uparrow}$,
$B_{2,\downarrow}$, $B_{1,\uparrow}$, $B_{1,\downarrow}$, $A_{2,\uparrow}$,
$A_{2,\downarrow}$ as
 $V=\left(\begin{smallmatrix} 0 & v^\dagger \\ v &
     0\end{smallmatrix}\right)$
with 
  \begin{equation}
    v = 
    \left(\begin{array}{c c c c}
      \tau p^* v_F & 0 & 0 & 2i\lambda_4 \delta_{\tau,1} \\
      0 & \tau  v_F p^* & 2i\lambda_4\delta_{\tau,-1} & 0 \\
      0 & 2i\lambda_4\delta_{\tau,-1} & \tau v_F p & 0 \\
      2i\lambda_4\delta_{\tau,1} & 0 & 0 & \tau v_F p 
    \end{array}\right)  ,
  \end{equation}
where $\delta_{\tau,\pm 1} = (1\pm\tau)/2$.

\begin{figure}[t]
    \includegraphics[width=\columnwidth]{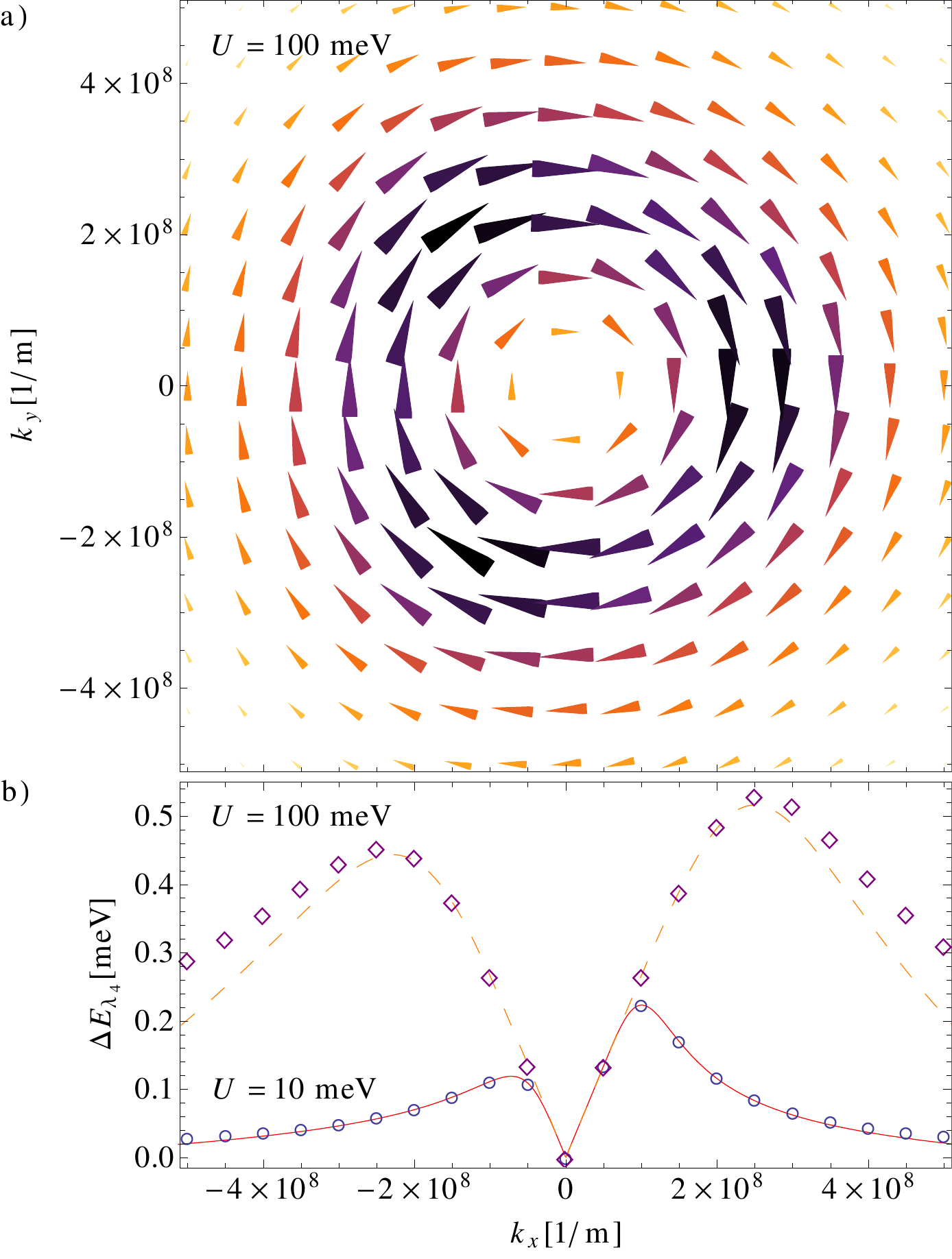}
  \caption{(a) Spin-orbit field $\mathbf{\Omega}_+(\mathbf{k})$ of electrons in bilayer graphene. 
(b)  Spin splitting $\Delta E$ where circles/diamonds refer to the
energy difference between the two 
electron-like low-energy bands $\Delta E=|E_{+,1}-E_{+,2}|$ obtained from a numerical diagonalization of the full Hamiltonian including $\lambda_4$-like SOI
and lines to the splitting of the spin-orbit field $\Delta E = \hbar
|\mathbf{\Omega_+(\mathbf{k})}|/2$ given 
by Eq.~(\ref{eq:OmegaLambda4e}).}
  \label{fig:deltaSOLambda4}
\end{figure}

We now perform the Schrieffer-Wolff transformation  
$\tilde H = e^SH e^{-S} = H_0 + \frac{1}{2}\left[S,V\right]$
where the anti-Hermitian matrix $S=-S^\dagger$ is determined by the
condition $V+[S,H_0]=0$, and where corrections of order
$(|p| v_F/\gamma_1)^3$ and $(\lambda_4/\gamma_1)^2$
have been neglected. 
The spin-independent part $\tilde H^0$ obtained from $\tilde H$ by setting
$\lambda_i=0$ for $i=1,\ldots,4$ 
reproduces the known form of the low-energy bands \cite{McCann2006}.
$E^0_\pm = \pm\left[\frac{U^2}{4}\left(1-2\kappa^2\right)^2+\gamma_1^2\kappa^2\left(
\kappa^2+\frac{v_3^2}{v_F^2}- 2\tau \kappa\frac{v_3}{v_F}\cos(3\phi)\right)\right]^{1/2}$,
where $\kappa = \hbar k v_F / \gamma_1$
and the (unnormalized) eigenstates 
\begin{equation}
  \psi_{\pm}^{\uparrow\downarrow}   = 
  \left(\begin{array}{c}
    \left|E^0_{\pm}\right|\pm U
      \left(1-2\kappa^2\right)
\\ 
\gamma_1\left(
        2\kappa^2-\tau\kappa  (v_3/v_F) e^{3i\phi}\right)
  \end{array}\right) \otimes
  \begin{cases}
    |\!\uparrow\rangle \\
    |\!\downarrow\rangle \\
  \end{cases} \;.
  \label{eq:psi0eff}
\end{equation}
in the absence of SOI. 
The spin-dependent part $H^{\lambda} = \tilde H-\tilde H^0$ can
be expressed in the eigenbasis Eq.~(\ref{eq:psi0eff}) of $\tilde H^0$,
\begin{equation}
 H^{\lambda} = 
\left( \begin{array}{c c}
H^{\lambda}_e & \Delta \\
\Delta^\dagger & H^{\lambda}_{h} 
\end{array}\right)
=
\left( \begin{array}{c c}
\frac{\hbar}{2}\mathbf{\Omega_+}\cdot \mathbf{s} & \Delta \\
\Delta^\dagger & \frac{\hbar}{2}\mathbf{\Omega_-}\cdot \mathbf{s}
\end{array}\right),
\end{equation}
with the electron (hole) effective spin-orbit field
\begin{equation}
  \mathbf{\Omega}_{\pm} \!=\! \frac{2\lambda_4U \kappa}{\hbar  E_\pm^0}
  \!\!\left[\!\left(1 \!-\! \kappa^2\right)
  \!\!\left(\!\!\!\begin{array}{c c c c}
   \sin\phi \\ -\cos\phi \\ 0 
  \end{array}\!\!\!\right)
   +\tau\kappa\frac{v_3}{v_F}
   \left(\!\!\begin{array}{c c c c}
    \sin 2\phi \\ \cos 2\phi \\ 0
   \end{array}\!\!\right)\!\!
   \right] \!\!.
  \label{eq:OmegaLambda4e}
\end{equation}
The spin-orbit field and splitting are shown in Fig.~\ref{fig:deltaSOLambda4} for two different values of the bias voltage, $U=0.1\,$eV and $U=0.01\,$eV. 
%In both cases $2\hbar|\vec\Omega_{\lambda_4}|$ gives a reasonable approximation even for wave vectors as large as $4\times10^8\,\text{m}^{-1}$.
For $\lambda_4\ll U$, the SOI-induced electron-hole coupling $\Delta$ can be neglected,
which is confirmed by our numerical analysis (see Fig. \ref{fig:deltaSOLambda4}b). 

As our next step, we derive the in- and out-of-plane spin relaxation
times originating from the presence of $\mathbf \Omega_{\mathbf k}\equiv\mathbf \Omega_{+}(\mathbf k)$ via the
D'yakonov-Perel' mechanism.  For concreteness, we restrict ourselves
to electrons.
Spin transport is modeled using a kinetic spin Bloch equation (KSBE),
i.e., 
a semiclassical rate equation for the spin distribution $\mathbf s_{\mathbf k}$ carried by an
ensemble of band electrons, an approach well known 
from semiconductor spintronics (see e.g. \cite{Fabian2007}).  
In the absence of external forces,
\begin{equation}
  \frac{\partial \mathbf s_{\mathbf k}}{\partial t} 
- \mathbf\Omega_{\mathbf k} \times \mathbf s_{\mathbf k}
%+\frac{1}{\hbar}\left(\mathbf F_{\mathbf k}\cdot\frac{\partial }{\partial
%    \mathbf k}\right) \mathbf s_{\mathbf k}
+ \frac{\partial\mathbf s_{\mathbf k}}{\partial \mathbf r}\cdot \mathbf v_{\mathbf k} 
 = \int\!\!\frac{d^2 k'}{(2\pi)^2} \!\left(W_{\mathbf k',\mathbf
    k}\mathbf s_{\mathbf k'} -  W_{\mathbf k,\mathbf k'}\mathbf s_{\mathbf k}\right) \!.
  \label{eq:ksbe}
\end{equation}
For the purpose of extracting the spin coherence times, it suffices to
consider the simplified scenario of a homogeneous spin distribution.
Furthermore, we restrict our calculation to elastic, i.e., energy
conserving, scattering and focus on the spin of the charge carriers at
the Fermi surface, which essentially corresponds to a zero-temperature
estimate. 
We consider Fermi energies $E_F$ much smaller than energy separation
of the split-off bands, but larger than 
the $|\mathbf k|=0$ offset of the low energy bands, i.e.\ $\gamma_1\gg E_F>U/2$. 
In this case there is a single connected Fermi surface near each of
the two valleys K and K' and we can employ
our effective low-energy theory with the spin-orbit field Eq.~(\ref{eq:OmegaLambda4e}).

At low energies, the energy bands experience a non-negligible
anisotropy due to the trigonal warping introduced by the interlayer
velocity $v_3$, which substantially complicates solving the KSBE. 
However, the corresponding effect on spin-relaxation is in most cases relatively small
which allows us to begin with a $v_3=0$ estimate, subsequently include
$v_3$ to first order, and finally compare our analytical results to a
numerical calculation taking trigonal warping fully into account.
A description of the last two steps as well as a discussion of the 
different results can be found in the Appendix. 
Here we only discuss the $v_3=0$ estimate.
Deviations from this result are comparably small and occur
predominantly where the Fermi energy is low and very close to $U/2$.

\begin{figure}[t]
  \begin{center}
    \includegraphics[width=0.48\textwidth]{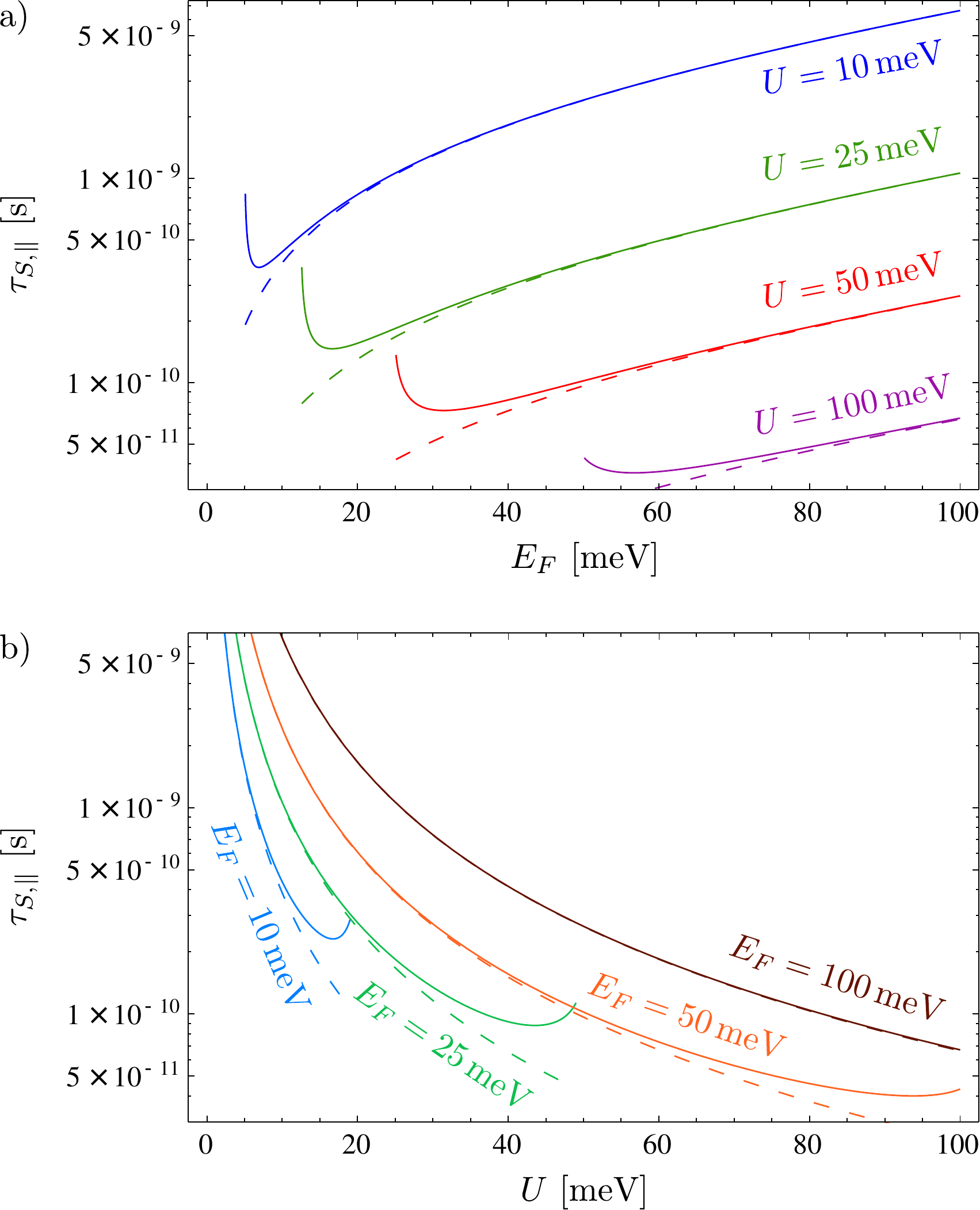}
  \end{center}
  \caption{In-plane spin relaxation time $\tau_{S,\parallel}$ in
    bilayer graphene with $\lambda_4$-type SOI in the isotropic limit
    ($v_3=0$), (a) as a function of the Fermi energy $E_F$ for
    different bias voltages $U$, and (b) as a
    function of the bias voltage $U$ at a constant Fermi energy 
   $E_F=0.06\,$eV. In both plots the momentum relaxation time is
   chosen to be $\tau_p=10^{-13}\,$s. 
   Dashed lines correspond to the large Fermi energy approximation 
   given in Eq.~(\ref{eq:tauSapprox}).}
  \label{fig:blgtauSofEF}
\end{figure}

In the isotropic limit $v_3=0$, the spin-orbit fields
$\mathbf\Omega_\pm(\mathbf k)$ given for electrons in Eq.~(\ref{eq:OmegaLambda4e})
and the band energies $E_\pm^0$ simplify considerably.
In particular, the magnitude of the spin-orbit field becomes isotropic
in this case, $|\mathbf\Omega_\pm(\mathbf k)|=\Omega_\pm(k)$, 
and is therefore constant if we consider electrons at the Fermi level
$|\mathbf p|=p_F=\hbar k_F$.
Moreover, in this limit the spin-orbit field becomes independent of the valley.
We can thus simply parameterize the spin
distribution in both valleys by the same angle $\phi$. 
In other words, it (formally) does 
not matter if the quasiparticle carrying the spin is located at K or K'. 
For elastic and symmetric scattering the scattering rates in Eq.~(\ref{eq:ksbe}) are of the form 
$W_{\mathbf k,\mathbf k'}=W_{\mathbf k',\mathbf k}= W(\phi-\phi')2\pi\hbar v_F\delta(E_{\mathbf k}-E_{\mathbf k'})$.
In the isotropic limit the collision integral only needs to be taken
over a circle of radius $|\mathbf k|=k_F$ and the KSBE 
given by Eq.~(\ref{eq:ksbe}) reduces to
\begin{equation}
  \frac{\partial \mathbf s_{\mathbf k}}{\partial t}-\mathbf\Omega_\mathbf{k}\times\mathbf s_{\mathbf k} =-\int\limits_0^{2\pi}\frac{d \phi'}{2\pi}\,W(\phi-\phi')\left(\mathbf s_{\mathbf k}-\mathbf s_{\mathbf k'}\right) \,.
  \label{eq:ksbeiso}
\end{equation}
In order to solve Eq.~(\ref{eq:ksbeiso}) we first decompose the spin
distribution function into an average $\mathbf s_0$ over the Fermi
surface, which is independent of the angle $\phi$, and the remaining
deviation $\Delta \mathbf s_{\mathbf k}$, describing the angular dependence,
 \begin{equation}
  \mathbf s_{\mathbf k} = \mathbf s_0 + \Delta \mathbf s_{\mathbf k} ,
\quad \quad
\mathbf s_0 \equiv \langle\mathbf s_{\mathbf k}\rangle \equiv \int\limits_0^{2\pi}\frac{d \phi}{2\pi} \mathbf s_{\mathbf k},
  \label{eq:decomp}
  \end{equation}
where
Note that the experimentally observed spin relaxation refers to the decay of the total spin of the charge carriers at the Fermi surface, which is in turn given by the average spin polarization $\mathbf s_0$.
To obtain the time dependence of $\mathbf s_0$ we substitute
Eq.~(\ref{eq:decomp}) into the KSBE (\ref{eq:ksbeiso}) 
and take the average over the angle $\phi$,
\begin{equation}
  \frac{\partial \mathbf s_0}{\partial t} = \langle \mathbf\Omega_\mathbf{k}\times \Delta\mathbf s_{\mathbf k}\rangle .
  \label{eq:ds0dt}
\end{equation}
Note that both the spin-orbit field and the collision integral average to zero. The corresponding equation for the anisotropic part is
\begin{eqnarray}
  \frac{\partial \Delta\mathbf s_{\mathbf k}}{\partial t} &=& \mathbf \Omega_\mathbf{k}\times \mathbf s_0 - \int\limits_0^{2\pi}\frac{d \phi'}{2\pi}\,W(\phi-\phi')\left(\Delta\mathbf s_{\mathbf k}-\Delta\mathbf s_{\mathbf k'}\right) \nonumber\\
  &&+\,\mathbf\Omega_\mathbf{k}\times \Delta\mathbf s_{\mathbf k} - \langle \mathbf\Omega_\mathbf{k}\times \Delta\mathbf s_{\mathbf k}\rangle \,.
  \label{eq:ddeltasdt}
\end{eqnarray}
The two coupled differential equations (\ref{eq:ds0dt}) and (\ref{eq:ddeltasdt}) can be solved approximately in the strong scattering limit $|\mathbf\Omega_{\mathbf k}|\tau_p\ll 1$, where $\tau_p$ denotes the momentum relaxation time. 
In this limit the combination of fast momentum scattering and slow spin precession implies that the 
deviation $\Delta\mathbf s_{\mathbf k}$ reaches a quasi-stationary
state $\mathbf \Delta s_{\mathbf k}^{\,\text{st}}$ when
$\partial\Delta\mathbf s_{\mathbf k}/\partial t\approx 0$,
which is then followed by a slow decay of the isotropic spin
polarization $\mathbf s_0$ \cite{Fabian2007}.
Since momentum relaxation is usually very fast on the time scale of the observation length, $\Delta t_\text{obs}\gg \tau_p$,
the observed dynamics of the spin polarization $\mathbf s_0$ is
effectively the averaged quantity $\mathbf s_0^{\text{obs}}(t)=\int_{t-\Delta t_\text{obs}/2}^{t+\Delta t_\text{obs}/2}\mathbf s_0(t')d t'$. 
We can therefore neglect fast fluctuations occurring on the time scale
$\tau_p$ as long as they are uncorrelated for times much longer than $\tau_p$.
It can be shown that the last two terms of Eq.~ (\ref{eq:ddeltasdt})
only give rise to fluctuations of the spin distribution, which are uncorrelated on a time scale $\gg \tau_p$.
Neglecting the last two terms of Eq.~(\ref{eq:ddeltasdt}) the steady state condition becomes
\begin{equation}
  \mathbf\Omega_\mathbf{k}\times \mathbf s_0 = \int\limits_0^{2\pi} \frac{d\phi'}{2\pi}W(\phi-\phi')\left(\Delta\mathbf s_{\mathbf k}^{\,\text{st}}-\Delta\mathbf s_{\mathbf k'}^{\,\text{st}}\right) \;.
  \label{eq:steadycond}
\end{equation}
This equation can be solved using the following ansatz,
\begin{equation}
  \Delta \mathbf s_{\mathbf k}^\text{\,st}  =  \tau^*
  \left[\mathbf\Omega_{\mathbf k}\times \mathbf s_0 \right],
  \label{eq:ansatzdeltas}
\end{equation}
where we still need to determine the time $\tau^*$. 
After substituting Eq.~(\ref{eq:ansatzdeltas}) into
Eq.~(\ref{eq:steadycond}), the integral can be treated by expanding 
the scattering rates $W(\phi-\phi')$ in polar harmonics. 
We find that Eq.~(\ref{eq:steadycond}) has the solution
\begin{equation}
  \frac{1}{\tau^*}  = \int\limits_0^{2\pi} \,\frac{d \theta}{2\pi} W(\theta) (1-\cos\theta),
  \label{eq:taustar}
\end{equation}
and therefore $\tau^*$ can be identified with the momentum relaxation
time, $\tau^*=\tau_p$. 
Having solved Eq.~(\ref{eq:steadycond}), we substitute the steady
state solution Eq.~(\ref{eq:ansatzdeltas}) 
with Eq.~(\ref{eq:taustar}) into the equation of motion of the total
spin polarization $\mathbf s_0$, Eq.~(\ref{eq:ds0dt}), and find
an exponential decay law,
\begin{equation}
  \frac{\partial\mathbf s_0^{\,\text{st}}}{\partial t} 
% = -|\mathbf\Omega_{\lambda_4}^0(k_F,\phi)|^2\tau_p 
%  \left(\begin{array}{c c c c}
%    \frac{1}{2} & 0 & 0 \\ 0 & \frac{1}{2} & 0 \\ 0 & 0 & 1
%  \end{array}\right)
  %\mathbf s_0^{\,\text{st}} 
=   \left(\begin{array}{c c c c}
    1/\tau_{S} & 0 & 0 \\
    0 & 1/\tau_{S} & 0 \\
    0 & 0 & 2/\tau_{S} \\
  \end{array}\right) \mathbf s_0^\text{\,st} ,
  \label{eq:s0decay}
\end{equation}
with the longitudinal spin-decoherence time
\begin{equation}
  \frac{1}{\tau_{S}} \equiv  \frac{1}{\tau_{S,\parallel} }
       =       \frac{2\lambda_4^2}{\hbar^2}\frac{U^2}{E_F^2}\kappa_F^2 
\left(1-\kappa_F^2\right)^2 \tau_p 
  \label{eq:tauSL4v30} ,
\end{equation}
where
\begin{equation}
  \kappa_F^2 = \frac{p_F^2v_F^2}{\gamma_1^2} = \frac{U^2+\sqrt{4E_F^2(U^2+\gamma_1^2)-U^2\gamma_1^2}}{2(U^2+\gamma_1^2)}
  \label{eq:pFv30}
\end{equation}
is found by solving $E_F=E_+^0|_{v_3=0}$.
The transverse spin relaxation time is simply $\tau_{S,\perp} = \tau_{S,\parallel}/2$.
Combining Eqs.~(\ref{eq:tauSL4v30}) and (\ref{eq:pFv30}), we obtain
the spin relaxation time as a function of the 
Fermi energy $E_F$ and the bias voltage $U$. 
As shown in Fig.~\ref{fig:blgtauSofEF}, %%%%GB
the spin relaxation time is very sensitive to both $E_F$ and $U$. 
For a constant $U$ and sufficiently large $E_F$ the spin relaxation
time increases as a function of $E_F$ and can be approximated by
\footnote{The result corresponds to a second order Taylor expansion in $U$. Note there are however two energy scales $U/\gamma_1$ and $\frac{U/2}{E_F}$.}
\begin{equation}
  \frac{1}{\tau_{S,\parallel}^0} = \frac{1}{2\tau_{S,\perp}^0} \approx \frac{2\lambda_4^2}{\hbar^2} \frac{(\gamma_1-E_F)^2U^2}{E_F\gamma_1^3}\tau_p \,.
  \label{eq:tauSapprox}
\end{equation}

The typical D'yakonov-Perel' relation $1/\tau_S\propto\tau_p$ has already been observed in two different experiments \cite{Yang2011} and \cite{Han2011}.
As pointed out in the previous discussion, the calculated relaxation
rates are very sensitive to a number of parameters: 
the Fermi energy $E_F$, the bias voltage $U$, and the SOI strength
$\lambda_4$.  
Unfortunately, these are not easily accessible experimentally. 
Thus, we will have to rely on some rough 
estimates in order to compare the experimental values of $\tau_S$ with the obtained theoretical results.

In \cite{Yang2011}, the spin relaxation time has been measured for for
a range of mobilities from $300\,\text{cm}^2/\text{Vs}$ to
$2000\,\text{cm}^2/\text{Vs}$ 
at room temperature and a range from 
$700\,\text{cm}^2/\text{Vs}$ to $3800\,\text{cm}^2/\text{Vs}$ at 5 K; 
both at a fixed carrier density $n_e=1.5\cdot
10^{12}\,\text{cm}^{-2}$. 
We can roughly estimate the Fermi energy using the parabolic
approximation $E(k)\approx \hbar^2 k^2/2m^*$, 
where $m^*=\gamma_1/2v_F^2$ is the effective mass \cite{McCann2006}. 
Integrating the density of states $D(E)$, which is constant within
this approximation, one obtains a carrier density 
of $n_e=m^* E_F/\pi\hbar^2$.  
The estimated Fermi energy for the experimental carrier density
$n_e=1.5\cdot 10^{12}\,\text{cm}^{-2}$ is $E_F=67\,$meV. 
Again, in the effective mass approximation we can estimate the
momentum relaxation 
time from $\tau_p=m^*\mu/e$ \cite{Yang2011}.
Assuming that the bias offset $U/2$ at $k=0$ is well below the Fermi
energy (which is also necessary for the parabolic approximation) 
we can use the approximate spin relaxation rate given by equation (\ref{eq:tauSapprox}),
\begin{align}
  \frac{1}{\tau_{S,\parallel}^0} \approx \frac{2\lambda_4^2}{\hbar^2} \frac{(\gamma_1-E_F)^2U^2}{E_F\gamma_1^3}\tau_p \overset{\tau_p\approx m^*\mu/e}{\approx} \frac{\lambda_4^2}{\hbar^2}\frac{(\gamma_1-E_F)^2U^2}{e v_F E_F\gamma_1^2} \mu .
  \label{eq:tauSapproxrep}
\end{align}
The experimental results show a reasonable agreement with the model
estimate for a bias voltage of 50$\,$meV. 
The corresponding model prediction Eq.~(\ref{eq:tauSapproxrep}) is
$\tau_{S}\approx 0.5\,\text{ns}\frac{1000\,\text{cm}^2 /\text{Vs}}{\mu}$.

At 5$\,$K the bias voltage of 50$\,$meV is significantly %%%%GB
larger than the thermal energy ($k_B T\approx 0.5\,$meV) and the Fermi
energy of 67$\,$meV reasonably well above $U/2$. At room temperature
($k_B T\approx 25\,$meV) the thermal energy is comparable with both,
thus making the zero temperature estimate very approximate. 
Interestingly, the experimental data shows a stronger correlation 
$\tau_S\propto 1/\mu$ at room temperature.

%%%%%

In conclusion, we have calculated the spin relaxation time in bilayer
graphene in dependence of the Fermi energy and interlayer bias
potential.  These two parameters can be tuned independently with
top and back gates.  Using experimentally determined parameters
and making reasonable assumptions for the unknown values of $U$
and $E_F$, we obtain good agreement with the existing experiments.
We find a strong dependence of the spin relaxation time on externally 
applied fields that may have applications in field-controlled spin valve
devices.

%%%%%

\textit{Acknowledgments.}
This work has been financially supported by DFG within FOR 912
and the ESF EuroGraphene project CONGRAN.

%\newpage

\clearpage

\appendix

\begin{widetext}

\section{Additional spin-orbit fields}
In the main text we have focused on $\lambda_4$-type SOI and derived the corresponding spin-orbit field.
Analogous spin-orbit fields can however be derived for the omitted terms of $H_\text{SO}$, i.e.\ $\lambda_1$-, $\lambda_2$-, and $\lambda_3$-type SOI.
In lowest order of the spin-orbit coupling constants we can consider each of the above spin-orbit terms separately.
For each term we can derive an analytic expression of the respective spin-orbit field using the same recipe as in the case of $\lambda_4$-type SOI. 
In order to compute $\mathbf\Omega_{\lambda_i}$ (for $i=1,2,3$) we start with $H=H_\text{BLG}\otimes\mathbbm{1}_S+H_\text{SO}\restriction_{\lambda_j=0\;\text{for }j\neq i}$, i.e.\ we omit all terms in $H_\text{SO}$ except for the one involving $\lambda_i$. 
Via a Schrieffer-Wolff transformation we separate high and low energy bands arriving at a effective low-energy Hamiltonian $\tilde H$, where we again neglect terms of order $\left(pv_F/\gamma_1\right)^3$ or $\left(\lambda_i/\gamma_1\right)^2$ and higher. 
The resulting effective Hamiltonian $\tilde H$ can be split into a kinetic and a spin-dependent part.
In all three cases we recover the same spin-independent part part $\tilde H^0$ as previously for $\lambda_4$. 
The remaining spin-dependent part $H^{\lambda}= \tilde H-\tilde H^0$ is subsequently rotated into the Eigenbasis of $\tilde H^0$ as given by Eq.~(\ref{eq:psi0eff}). 
For a sufficiently a large bias  ($U\gg\lambda_i$), the electron-hole coupling $\Delta$
can be dropped. %%%%GB
Form the remaining $2\times 2$ blocks $H_{e/h}^{\lambda_i}$ %%%%GB
we obtain the respective spin-orbit field $\mathbf\Omega_{\pm}^{\lambda_i}$.

Below we report the resulting expressions for the approximate spin-orbit fields:
\begin{eqnarray}
  &\mathbf\Omega_\pm^{\lambda_1} = \frac{2\lambda_1}{\hbar}\frac{U}{E^0_\pm}\left[\frac{1}{2}-\kappa^2\left(1-2\kappa^2+2 \tau\kappa\frac{v_3}{v_F}\right)\right] 
  \begin{pmatrix}
    0 \\ 0 \\ 1
  \end{pmatrix} \,,&
  \label{eq:omegaL1} \\
  &\mathbf\Omega_\pm^{\lambda_2} = \frac{2\lambda_2}{\hbar}\frac{U}{E^0_\pm}\left[\frac{1}{2}-2\kappa^2\left(1-2\kappa^2\right)\right]
  \begin{pmatrix}
    0 \\ 0 \\ 1
  \end{pmatrix} \,,&
  \label{eq:omegaL2} \\
  &\mathbf\Omega_\pm^{\lambda_3} = \pm\frac{2\lambda_3}{\hbar}\frac{U}{\gamma_1}2\kappa
  \begin{pmatrix}
    \sin\phi \\ \cos\phi \\ 0
  \end{pmatrix} \,.&
  \label{eq:omegaL3} 
\end{eqnarray}
Note that both $\mathbf\Omega_\pm^{\lambda_1}$ and
$\mathbf\Omega_\pm^{\lambda_2}$ are out-of-plane effective magnetic
fields, which in the isotropic limit ($v_3=0$) are independent of the electron %%%%GB
momentum. Similar to $\mathbf\Omega_\pm^{\lambda_4}$,
$\mathbf\Omega_\pm^{\lambda_3}$ is an in-plane effective field, which
changes its direction depending on the angle of the electrons momentum
$\phi$. In contrast to $\mathbf\Omega_\pm^{\lambda_4}$ it is however
not proportional to $U/E_\pm^0$, but instead to $U/\gamma_1$. In other
words
$|\mathbf\Omega_\pm^{\lambda_3}|/|\mathbf\Omega_\pm^{\lambda_4}|\propto
(\lambda_3/\lambda_4)(E_\pm^0/\gamma_1)$, which in the range of the
low-energy theory makes it small even %%%%GB
if $\lambda_3$ and $\lambda_4$ were comparable.

\section{Spin-relaxation - a first order estimate including trigonal warping}
In the analytic derivation of the in- and out-of-plane spin relaxation
rates given in the main text we have 
neglected the anisotropy of the band structure.  
In the case of a finite trigonal warping ($v_3\neq 0$) the length of
the Fermi wave vector is no longer constant on the Fermi surface. 
Moreover, the density of states at the Fermi level is no longer constant. 
Solving the general scattering integral, which previously used to be a
simple integral over the angle, now becomes a more complicated task.  
In order to obtain a first estimate of the effect of trigonal warping on the spin relaxation time we instead choose a much simpler approach. 
Namely, we use the (momentum) relaxation time approximation of the KSBE. 
Here the scattering integral is replaced by a single parameter, the momentum relaxation time:
\begin{equation}
 \frac{\partial \mathbf s_{\mathbf k}}{\partial
    t}-\mathbf\Omega_{\mathbf k}\times\mathbf s_{\mathbf k}
  =-\tau_p\left(\mathbf s_{\mathbf k}-\langle\mathbf s_{\mathbf
      k}\rangle\right) \,,
  \label{eq:ksberelax}
\end{equation}
where $\langle \cdot\rangle$ denotes the average over the Fermi surface and $E(\mathbf k)=E_F$, as we assume elastic scattering and only consider electrons at the Fermi level. 
Although we may not be able to calculate the average $\langle
\cdot\rangle$ analytically,  %%%%GB
we can use a semi-numerical approach. 
Therefore, we again decompose the spin distribution function into its average and its $\mathbf k$-dependent deviation:
\begin{equation}
  \mathbf s_{\mathbf k} = \langle\mathbf s_{\mathbf k}\rangle + \Delta\mathbf s_{\mathbf k}
  \label{eq:decomposev3}
\end{equation}
Neglecting the fluctuation term 
$\mathbf\Omega_{\mathbf k}\times \Delta\mathbf s_{\mathbf k} - \langle \mathbf\Omega_{\mathbf k}\times \Delta\mathbf s_{\mathbf k}\rangle$,
the kinetic spin Bloch equation simplifies to
\begin{equation}
 \frac{\partial \langle\mathbf s_{\mathbf k}\rangle}{\partial t} = \langle \mathbf\Omega_{\mathbf k}\times \Delta\mathbf s_{\mathbf k}\rangle  \qquad\text{and}\qquad %%%%MD subscript
  \frac{\partial \Delta\mathbf s_{\mathbf k}}{\partial t} = \mathbf \Omega_{\mathbf k}\times \langle\mathbf s_{\mathbf k}\rangle -\tau_p\Delta\mathbf s_{\mathbf k} \,. %%%%MD subscript
  \label{eq:decompksbe}
\end{equation}
The steady-state solution of the $\mathbf k$-dependent part is readily
given by 
$\Delta\mathbf s_{\mathbf k}^\text{\;st} \equiv 
\tau_p\left(\mathbf\Omega_{\mathbf k}\times\langle\mathbf s_{\mathbf
    k}\rangle\right)$.  %%%%GB
%%%%MD subscript
%
For the corresponding time evolution of the average spin we find %%%%GB
\begin{equation}
  \frac{\partial \langle\mathbf s_{\mathbf
      k}\rangle^\text{st}}{\partial t} 
= -\tau_p\left\langle \Omega^2_{\mathbf k}\,\langle\mathbf s_{\mathbf k}\rangle^\text{st}
-\left(\mathbf\Omega_{\mathbf k}\cdot\langle\mathbf s_{\mathbf
    k}\rangle^\text{st}\right) \mathbf\Omega_{\mathbf k}\right\rangle \,. %%%%MD subscript
  \label{eq:dssteady}
\end{equation}
The part of the right hand side that is proportional to  %%%%GB
$\langle\mathbf s_{\mathbf k}\rangle^\text{st}$ leads to the first order
estimate of the spin relaxation time including trigonal warping,
\begin{equation}
  \frac{1}{\tau_{S,i}^1} = -\tau_p \left\langle\Omega^2_{\mathbf k} - \Omega^2_{\mathbf k,i}\right\rangle \qquad\text{for}\qquad i=x,y,z\,. %%%%MD subscript
  \label{eq:tauS1}
\end{equation}
Since the out-of-plane component $\Omega_{\mathbf k,z}$ simply vanishes and the two in-plane components are of the same average amplitude in both valleys, we can immediately recover that there is still only two different spin relaxation times ($\tau_{S,\parallel}^1$ and $\tau_{S,\perp}^1$).

In order to numerically calculate Eq.~(\ref{eq:tauS1}),
we need to explicitly calculate the average over the Fermi surface.  %%%%GB
In the case of $\gamma_1\gg E_F>U/2$ this can be achieved by a numerical inversion of the low-energy dispersion relation. 
Inverting $E_+^0(k,\phi)$ at a discrete number of angles and using a standard interpolating function, we obtain $k_F(\phi)$, i.e.\ the amplitude of the Fermi vector as a function of its angle. 
The average over the Fermi surface can be expressed in terms of a single integral over the angle:
\begin{equation}
  \langle f(k,\phi)\rangle = \frac{1}{Z}\int\limits_0^{2\pi} d\phi \,D(\phi) f\left[k_F(\phi),\phi\right] \,,
  \label{eq:Fermiaverage}
\end{equation}
where
\begin{equation}
  D(\phi) = \frac{1}{2\pi^2}\frac{\sqrt{[k_F(\phi)]^2+[\partial_\phi k_F(\phi)]^2}}{\sqrt{(\partial_k E_\text{eff}^0(k,\phi)\restriction_{k=k_F(\phi)})^2+(\frac{1}{k}\partial_\phi E_\text{eff}^0(k,\phi)\restriction_{k=k_F(\phi)})^2}}
  \label{eq:Dofphi}
\end{equation}
is the respective density of states and $Z=\langle
D(\phi)\rangle$. The above density of states along the anisotropic
Fermi surface can be derived form a coordinate transformation into
local coordinates $k_\parallel$ and $k_\perp$, pointing along and
perpendicular to the Fermi surface.

\section{A numerical model of spin relaxation}
To check the approximations we have employed when solving the KSBE, we
also consider a simple numerical model  
that simulates the concept of the D'yakonov-Perel' mechanism. %%%%GB 
We therefore sample the spin evolution of an ensemble of electrons at the Fermi level. 
The diffusive (real space) motion of the ensemble is modeled by a random k-space walk of each electron. 
A homogeneous (or averaged) spin-orbit interaction is represented by a
$\mathbf k$-dependent spin-orbit field 
$\mathbf\Omega_{\mathbf k}$, %%%%MD subscript
which in turn acts on the spin of each electron. 
Following the semiclassical approximation we assign each electron-like
quasiparticle a wave 
vector $\mathbf k$ (relative to one of the Dirac-points) and a spin $\mathbf S$. 
Their dynamics are governed by semiclassical equations of motion,
i.e.\ in absence of external forces, unless the electron is being
scattered, $\mathbf k$ is simply constant an $\mathbf S$ evolves
according to 
$\partial\mathbf{S}/\partial t=\mathbf\Omega_{\mathbf k}\times\mathbf S$. %%%%GB

Momentum scattering on the other hand is modeled by a homogeneous
scattering rate $W(\mathbf k,\mathbf k')$, 
representing the rate at which electrons in state $\mathbf k$ scatter
into the state $\mathbf k'$. 
Scattering is assumed to be elastic and spin conserving. 
For a simple model we consider scatterers to be represented by
Gaussian model potentials of width 
$R$, i.e., $V(\mathbf r)\equiv V_0\exp(-r^2/2R^2)$. %%%%GB
%A calculation of the resulting chiral scattering cross section can be
%found in Ref. \cite{PhysRevB.77.115433}.
Here we study small scatterers, where the spread of the potential is
still larger than the lattice constant, but much smaller than the inverse of the wave vector amplitude: $a\ll R\ll 1/k$.
In this limit the explicit $|\mathbf k|$-dependence can be neglected and the scattering cross section simplifies to $d\sigma/d\theta\propto \cos^2\theta$, where $\theta$ is the scattering angle.
The remaining dependence $\cos^2\theta$ is the signature of the  Berry
phase of the quasi particles. %%%%GB
Note that $R\ll 1/k$, where $k\ll K$, also implies that intervalley scattering can be neglected.
According to Fermi's golden rule the scattering rate form $\mathbf k$ to $\mathbf k'$ is proportional to the density of states at the outgoing momentum $\hbar\mathbf k'$.
If we again focus on Fermi energies sufficiently larger than the $\mathbf k=0$ bias offset $U/2$, each electron wave vector $\mathbf k$ can be parameterized by its angle $\phi$, where $|\mathbf k|=k_F(\phi)$ (see previous section).
This suggests the following expression for the scattering rate for small scatterers in bilayer graphene with finite trigonal warping,
\begin{align}
  W(\phi,\phi') \equiv  \frac{1}{\tau_{sc}} \,\frac{1}{Z}\, \cos^2(\phi-\phi') \,D(\phi')\,,
  \label{eq:dsigmadthetaBLG}
\end{align}
where $Z=\int d\phi'd\phi \cos^2(\phi-\phi') \,D(\phi')$ is the normalization, $1/\tau_{sc}$ the total scattering rate and $D(\phi')$ the angular dependent density of states as given by Eq.~(\ref{eq:Dofphi}).
All of the numerical results presented in Fig. \ref{fig:A1} are
calculated using this approximation. Note that in the isotropic limit
$D(\phi')=\text{const.}$ implies that the momentum relaxation time
$\tau_p$ (see Eq.~\ref{eq:taustar}) is equal to the mean
scattering time $\tau_{sc}$.

\section{Fermi energy and bias voltage dependence - a comparison with numerics}
Fig. \ref{fig:A1} shows the numerical spin relaxation in comparison
with the two estimates for a range of bias voltages from 10$\,$meV to
100$\,$meV and different Fermi energies. 
As previously noted, all Fermi energies are chosen to be larger than
the $k=0$ offset given by the bias voltage ($E_F>U/2$).
The calculated examples demonstrate an excellent agreement between the
numerical data and the above first order estimate
$\tau_{S,\parallel}^1$ (\ref{eq:tauS1}) for all (b through f) but the
first example (a). In these cases even the zeroth order estimate
$\tau_{S,\parallel}^0$, where $\tau_p=\tau_{sc}$, is in comparably
good agreement with the numerical data. Noteworthy deviations only
occur in the cases (d and f), where the Fermi energy is very close to
the voltage offset ($E_F\simeq U/2+0.1\,$meV). %%%%GB
As shown in the corresponding insets these are exactly the cases where trigonal warping is most pronounced. (a) is the only example where both $\tau_{S,\parallel}^0$ and $\tau_{S,\parallel}^1$ deviate significantly from the numerical results. However, taking into account the extreme trigonal warping, both still provide a good order of magnitude estimate. Overall, the numerical data supports the sensitive dependence of the spin relaxation time on bias voltage and Fermi energy shown in Fig. \ref{fig:blgtauSofEF}. Notice that there is roughly two orders of magnitude difference between the spin relaxation times for $U=10\,$meV and $U=100\,$meV at $E_F=50.1\,$meV (see table \ref{tab:A1}).

Though not explicitly shown here, the numerical calculations indicate the same anisotropy factor of two between the relaxation times of the in- and out-of-plane spin polarization, which we derived analytically in the isotropic limit $v_3=0$.

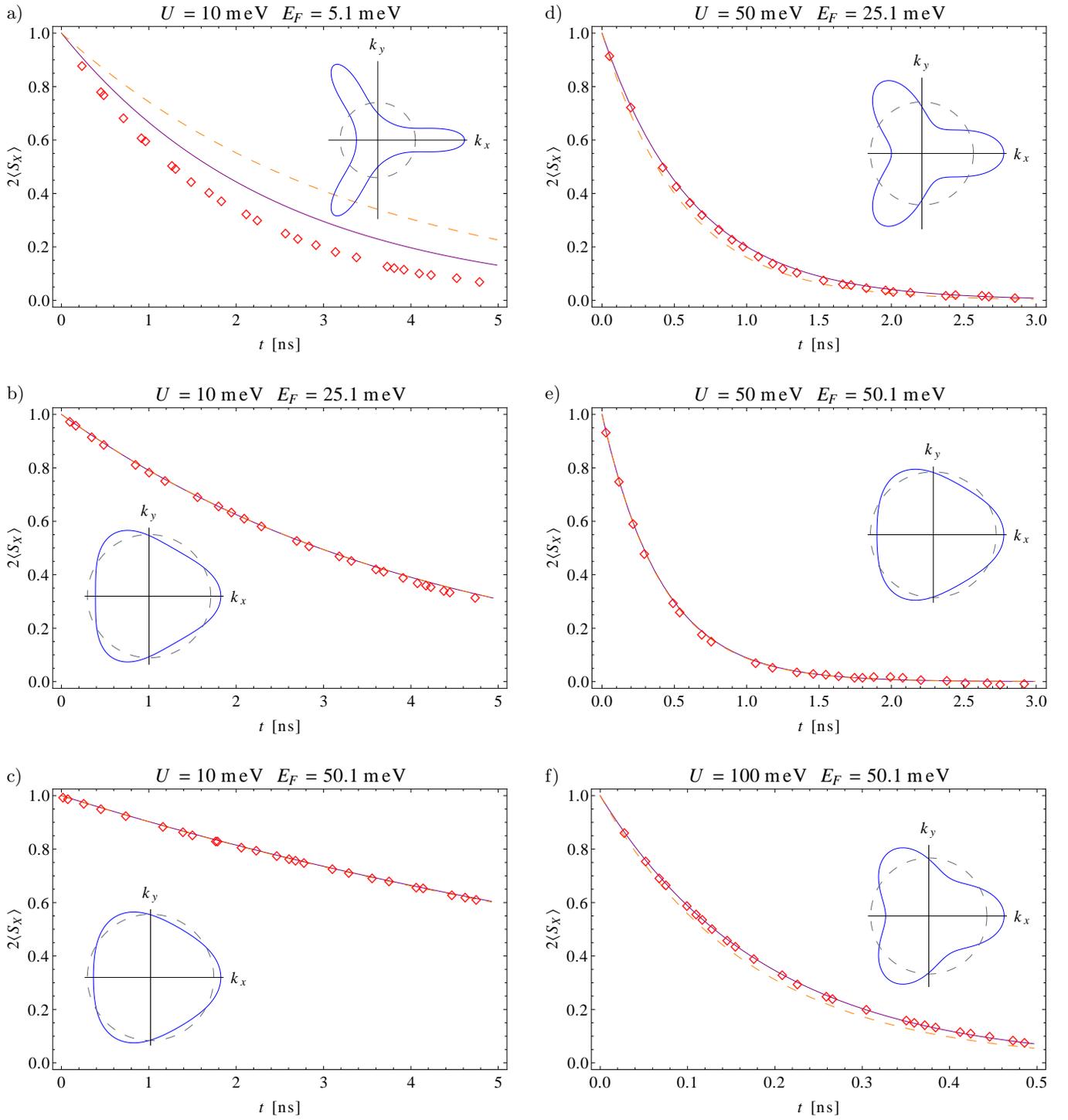
\begin{figure*}[htp]
  \begin{overpic}[width=\textwidth,tics=5,trim=14mm 0mm 12mm 0mm]{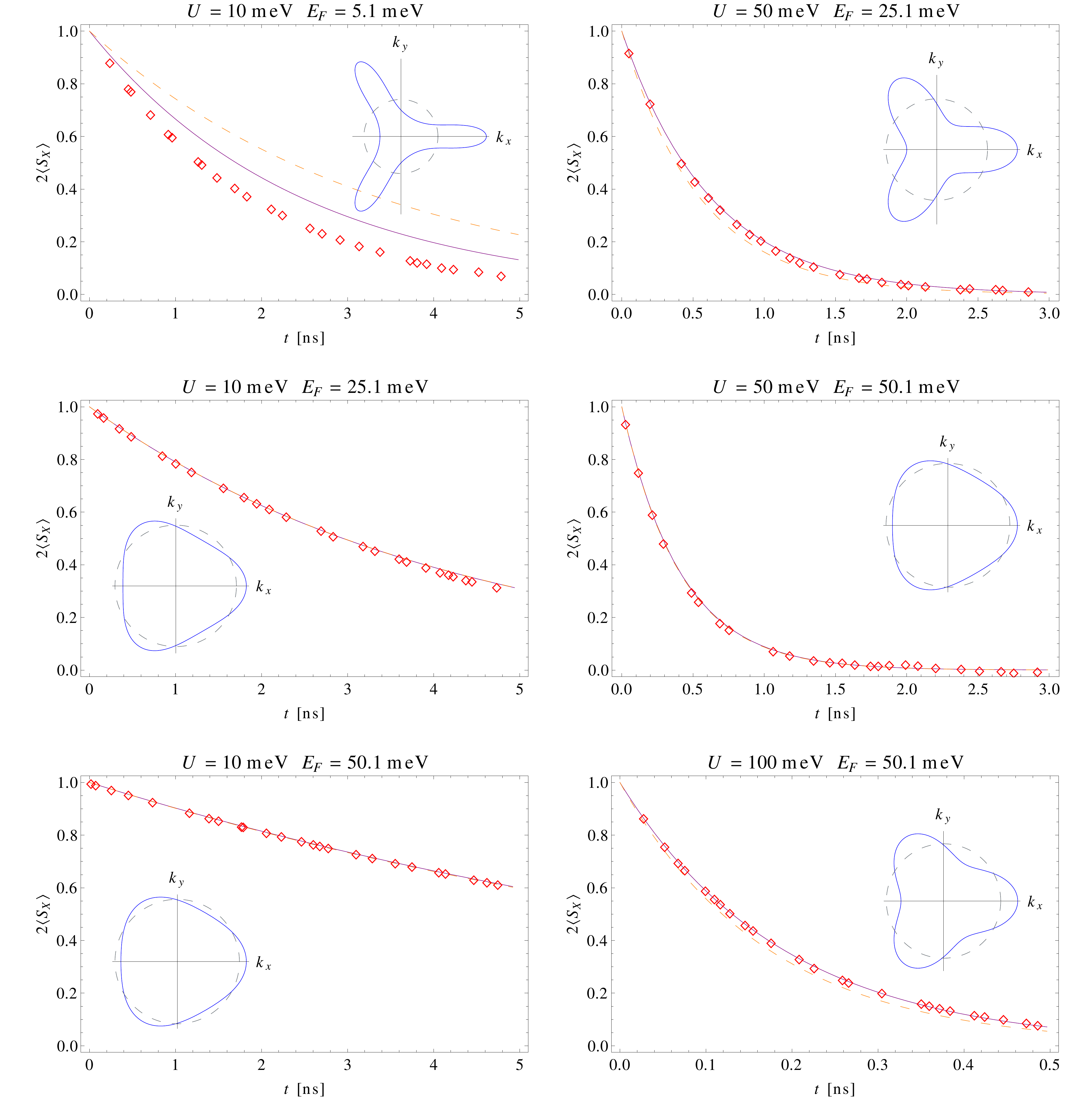} %%%%MD use option grid to see where to put the labels
    \put(0,98.5){\small a)}
    \put(48.3,98.5){\small d)}
    \put(0,64.3){\small b)}
    \put(48.3,64.3){\small e)}
    \put(0,30){\small c)}
    \put(48.3,30){\small f)}
  \end{overpic}
    \caption{(color online) In-plane spin relaxation in bilayer
      graphene with $\lambda_4$-type SOI. 
The plots show a comparison of the numerical data (diamonds) and the
zeroth (orange dashed) and first (continuous purple lines) order
estimates, $\tau_{S,\parallel}^0$ (\ref{eq:tauSL4v30}) and
$\tau_{S,\parallel}^1$ (\ref{eq:tauS1}), for different values of  the
interlayer bias $U$ and the Fermi energy $E_F$. All curves are
calculated in the limit of $a\ll R \ll 1/k_F$, using a mean scattering
time $\tau_\text{sc}=\tau_p=0.1\,$ps. 
The Insets show the trigonal warping of the Fermi surfaces. 
The corresponding spin relaxation times are listed in table \ref{tab:A1}.}
    \label{fig:A1}%%%%GB
\end{figure*}
  
\begin{table}
  \begin{ruledtabular}\begin{tabular}{ccccc}
  $U\,$[meV] & $E_F\,$[meV] & $\tau_0\,$[ns] & $\tau_1\,$[ns] &
  $\tau_{fit}\,$[ns] \\[0.5mm] \hline 
  10 & 5.1 & 3.36 & 2.46 & 1.99 \\
  & 25.1 & 4.26 & 4.25 & 4.20 \\
  & 50.1 & 9.69 & 9.78 & 9.80 \\   
  50 & 25.1 & 0.547 & 0.623 & 0.612 \\
  & 50.1 & 0.409 & 0.413 & 0.407 \\                       
  100 & 50.1 & 0.171 & 0.188 & 0.190 \\ 
  \end{tabular}\end{ruledtabular}
  \caption{In-plane spin relaxation time in bilayer graphene with
    $\lambda_4$-type SOI for different bias voltages 
$U$ and Fermi energies $E_F$. 
The table is a comparison of the zeroth and first order estimates, $\tau_{S,\parallel}^0$ (\ref{eq:tauSL4v30}) and $\tau_{S,\parallel}^1$ (\ref{eq:tauS1}), with the best-fit values for the numerical data shown in Fig. \ref{fig:A1}.}
  \label{tab:A1}  %%%%GB
\end{table}

\end{widetext}

\end{document}